\documentclass[twocolumn,prl,superscriptaddress]{revtex4-1}
\usepackage{amsmath}
\usepackage{graphicx}
\usepackage{bm}
\usepackage{color}
\usepackage{amsfonts}
\usepackage{dcolumn}

\begin{document}
\title{Origin of ultrastability in vapor-deposited glasses}

\author{Ludovic Berthier}
\affiliation{Laboratoire Charles Coulomb, UMR 5221, Universit\'e de Montpellier and CNRS, 
34095 Montpellier, France}
\author{Patrick Charbonneau}
\affiliation{Department of Chemistry, Duke University, Durham, NC 27708; \\ Department 
of Physics, Duke University, Durham, NC 27708}
\author{Elijah Flenner}
\affiliation{Department of Chemistry, Colorado State University, Fort Collins, CO 80523}
\author{Francesco Zamponi}
\affiliation{Laboratoire de physique th\'eorique, Ecole normale sup\'erieure, PSL
Research University, Sorbonne Universit\'es, UPMC Univ. Paris 06, CNRS, 75005 Paris, France}
\date{\today}

\begin{abstract}
Glass films created by vapor-depositing molecules onto a substrate 
can exhibit properties
similar to those of ordinary glasses aged for 
thousands of years. It
is believed that enhanced surface mobility 
is the mechanism that allows vapor deposition to 
create such exceptional glasses, but it is 
unclear how this {effect} is related to the final state of the film.
Here we use molecular dynamics simulations 
to model vapor deposition and an efficient Monte Carlo algorithm 
to determine the deposition rate needed to create ultra-stable glassy films.   
We obtain a scaling relation that
quantitatively captures the efficiency gain of vapor deposition over 
bulk annealing, and demonstrates
that surface relaxation plays the same role in the formation of 
vapor-deposited glasses as bulk relaxation does in 
ordinary glass formation. 
\end{abstract}

\maketitle

Compared to their liquid-cooled counterparts, vapor-deposited glasses often have 
a higher density \cite{Dalal2012a}, a higher kinetic stability \cite{Rodriguez2016,Swallen2007,Leon2010}, 
and a lower heat capacity \cite{Kearns2010}. This makes
them promising materials for a wide range of applications, such as 
drug delivery \cite{Hancock1997}, 
protective coatings \cite{Yu2013,Chu2012}, and lithography \cite{Neuber2014}.
Identifying the microscopic process that gives rise to these properties
is thus crucial to designing novel amorphous materials~\cite{today2016}.   
{Vapor deposition indeed} does not systematically result in glasses
with improved characteristics. It is observed that the substrate 
ought to be held at a specific temperature
(around 85\% of the glass transition 
temperature $T_g$ of the liquid~\cite{Swallen2007}) and that the deposition 
rate {must} be sufficiently slow~\cite{Chua2015} to get optimal 
films. A microscopic explanation for the optimality of $0.85 T_g$, 
and an 
estimate of 
what is a ``sufficiently slow'' deposition rate are, however, still lacking.
Moreover, while simulations and experiments 
have shown that vapor-deposited glasses may
lie lower in the potential energy landscape than liquid-cooled 
glasses~\cite{Swallen2007,Singh2011,Lyubimov2013,Lyubimov2015,Reid2016,Chua2015,starr2017},
and sometimes 
have the same structure as glasses of a comparable energy \cite{Reid2016},
it is not known whether vapor deposition can provide truly 
equilibrium 
configurations, especially below $T_g$. 

\begin{figure}[b]
\includegraphics[width=8.5cm]{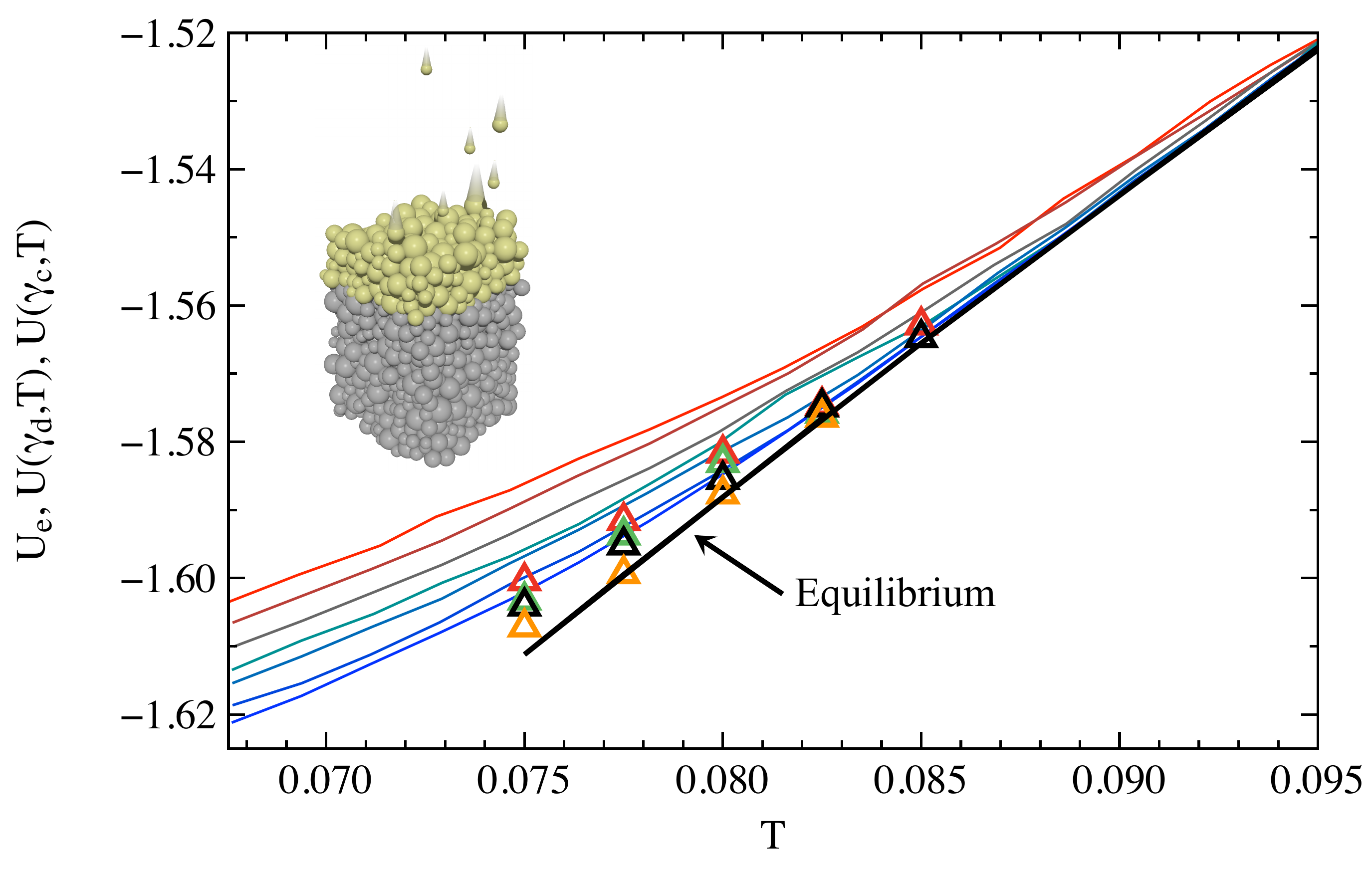}
\caption{\label{energy}
Average potential energy for ordinary liquid-cooled films (lines) for cooling
rates of $\gamma_c = 10^{-5}$, $5\times10^{-6}$, $2\times10^{-6}$, $10^{-6}$, $5\times10^{-7}$, $2\times10^{-7}$, 
and $10^{-7}$ listed from top to bottom. The open triangles are for vapor-deposited films {with} deposition rates
$\gamma_d=2.2\times10^{-3}$ (red), $7.3\times10^{-4}$ (green), $2.2\times10^{-4}$ (black), and $4.4\times10^{-5}$ (orange). 
The black line is the potential energy 
for the equilibrium supercooled liquid film. Inset: illustration of the 
vapor deposition with the growing film (red particles) 
and a temperature-controlled 
substrate (green particles). 
}
\end{figure}

Here we provide a quantitative test 
of the role of surface mobility in the creation of vapor deposited glasses. More specifically, we answer two key questions. 
{\it (i)} How much more efficient is vapor deposition than standard 
cooling in creating a glass or, more precisely, given a substrate temperature and a deposition rate, what is the effective cooling rate that would produce similar configurations?
{\it (ii)} What is the deposition rate needed to produce fully equilibrated
configurations?
Answering these questions is a challenging program 
that requires characterizing 
equilibrated films at temperatures sufficiently low for a large
difference between surface and bulk relaxation to have developed,
as well as measuring bulk and surface dynamics in a same material, over 
the same temperature 
range, and under the same thermodynamic conditions. 
We overcome these problems by using,
on a properly chosen polydisperse Lennard-Jones model, a swap Monte Carlo algorithm that efficiently samples the energy 
landscape at very low temperatures, speeding up equilibration by several orders of magnitude 
over standard molecular dynamics~\cite{Berthier2016,Ninarello2017}.
This allows us to 
compare free standing 
equilibrated 
films with those grown 
on an equilibrated substrate using an 
algorithm that closely mimics experimental vapor deposition,
and to independently
determine the low temperature equilibrium energy of the film
with no extrapolation. 

\begin{figure}
\includegraphics[width=8.5cm]{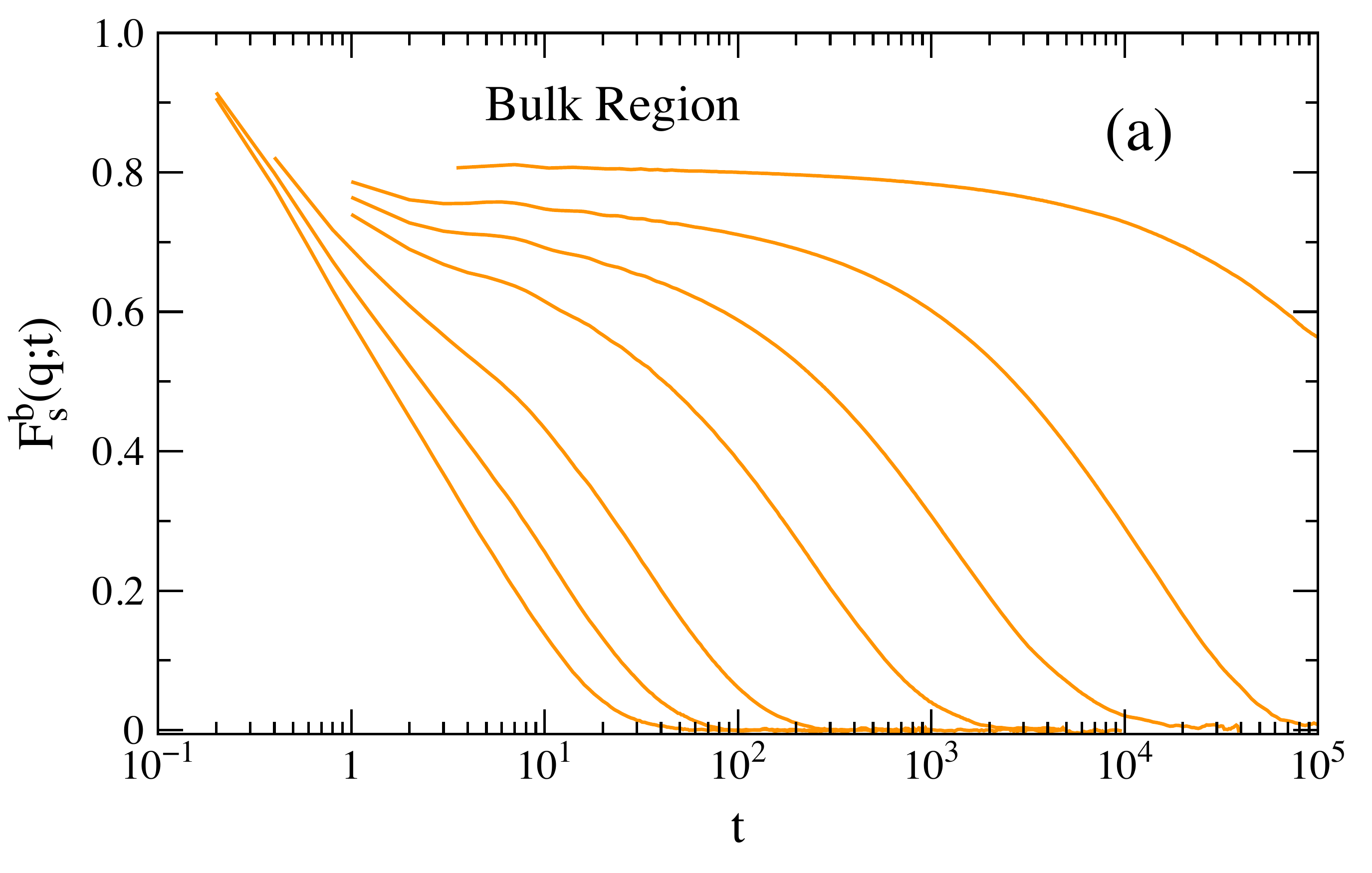}
\includegraphics[width=8.5cm]{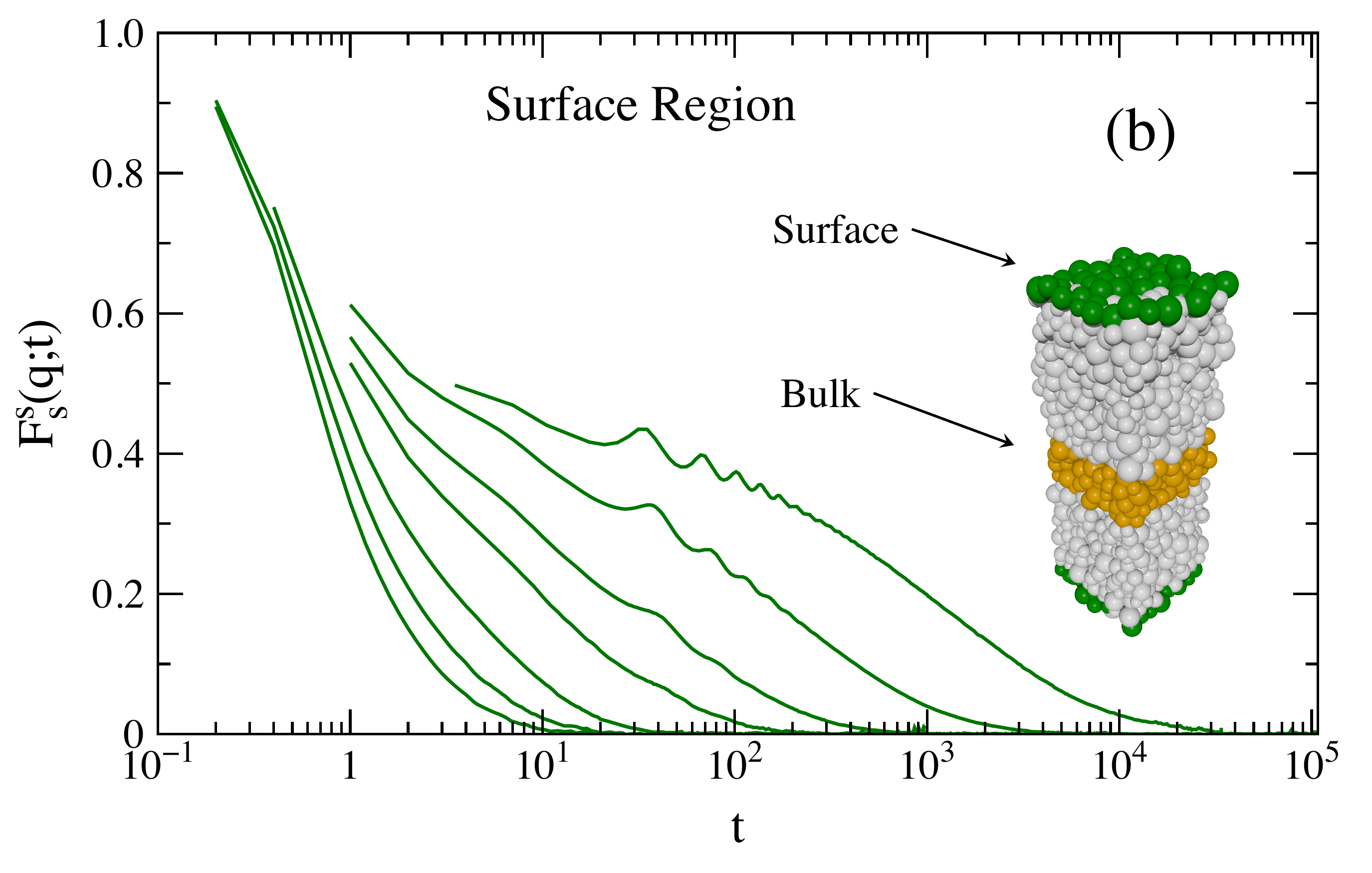}
\caption{\label{fs}
(a) Self-intermediate scattering function calculated for particles initially within the 
bulk region of an equilibrated supercooled liquid film, $F_s^b(\mathbf{q};t)$, for $T=0.12$, 0.11, 0.1, 0.09, 0.085, 0.08, and 0.075, from left to right.
(b) Self-intermediate scattering function calculated for particles initially at the film surface, $F_s^s(\mathbf{q};t)$, for the same temperatures. The inset illustrates the extent of the surface
and the bulk regions of an equilibrium freestanding film.}
\end{figure}

We simulate a polydisperse mixture of Lennard-Jones particles with interaction 
potential, $V(r_{ij}) = \epsilon \left[(\sigma_{ij}/r_{ij})^{12} - (\sigma_{ij}/r_{ij})^{6} \right]$,
which is truncated and shifted to zero at $2.5 \sigma_{ij}$.
The size parameter $\sigma_{ij}$ is given by a non-additive mixing rule 
$\sigma_{ij} = 0.5(\sigma_i + \sigma_j)(1 - \Delta  |\sigma_i - \sigma_j|)$, where $\Delta = 0.2$. The mixing parameter $\Delta$ is chosen to avoid separation of the particles into small and large components at low 
pressures and temperatures~\cite{Ninarello2017}. 
The particle size parameter $\sigma$ is chosen from 
the probability distribution 
$P(\sigma) = A/\sigma^3$ for $0.73 \le \sigma \le 1.62$ and
zero otherwise~\cite{Ninarello2017}. 
For each temperature we simulate five different realizations of the 
size distribution. Each particle has
the same mass $m$, the unit of energy is $\epsilon$, and the unit of length is the average of the particle diameters
$\sigma_0$. Our unit of time is $\sqrt{\sigma_0^2 m/\epsilon}$. The 
freestanding films have box lengths of $11 \sigma_0$ in the periodic $x$ and $y$ directions. 
The box length in the $z$ direction is 120$\sigma_0$. We simulate systems of $N = 4000$ particles, which results in
films of around 30$\sigma_0$ along the $z$ direction. 

To create equilibrium free standing 
films we used a Monte Carlo swap algorithm. The 
algorithm consists of two Monte Carlo moves, a standard attempted displacement move
and an attempted swap move. For the attempted displacement move we use trial positions 
within a cube of side $d$, where $d$ is adjusted for each temperature so that the acceptance 
rate lies between 0.3 to 0.35. For the attempted swap move we consider exchanging the size 
of two particles chosen at random. The move type is chosen at random, with 
20\% of the moves being attempted swaps. 

For the vapor-deposited films we first create an equilibrated substrate of the same system with 
$N/2 = 2000$ particles in a simulation box of the same dimensions as for the free-standing film. We then introduce $N/2$ particles with $x$ and $y$ components of the velocity randomly chosen within the square of side 
0.02. The $z$ plane from which particles are introduced moves at a constant velocity in order to remain about $40 \sigma_0$ 
above the surface of the substrate. The velocity of the deposited 
particles in the $z$ direction has a magnitude
corresponding to a kinetic temperature of $T=0.1$
(the onset temperature of slow dynamics), directed towards the substrate. The total momentum of the whole 
system is set to zero at every time step to reduce the drift of the center of mass. 
The substrate is simulated using a Nos\'e-Hoover constant $NVT$ algorithm, and the vapor-deposited
particles are simulated using a constant $NVE$ algorithm. After all the particles are deposited onto the
substrate, the simulation is run for $t=100$ and the following $t=1000$ is used to calculate the
average energy for that deposition rate. In all cases the energy changes little over the
averaging window.

Figure~\ref{energy} compares the temperature evolution of the
average potential energy, $U$, measured in the exact same film geometry 
{and} at the same temperature $T$ {for} three different protocols.
Vapor-deposited films at deposition rate $\gamma_d = dz/dt$ and substrate temperature $T$
have energy $U(\gamma_d,T)$ (triangles),
ordinarily-cooled films at cooling rate $\gamma_c = dT/dt$ 
have energy $U(\gamma_c,T)$ (colored lines), and 
films equilibrated using swap Monte Carlo have 
energy $U_e(T)$ (black line).
{In agreement with prior experiments and 
simulations~\cite{Lyubimov2015,Reid2016,starr2017}, 
the results show that}
vapor deposition equilibrates our model more efficiently than liquid 
cooling for a comparable preparation time, especially at low temperatures.
The average energy obtained at slower cooling rates 
indeed deviates from equilibrium below $T \approx 0.085$, 
while the films grown by vapor deposition have energies much  
closer to equilibrium. For our slowest deposition rate, the average energy
remains equal to the equilibrium energy for all but the lowest temperature 
studied, $T=0.075$. 

We are {now} in the position to answer 
the key questions raised above. 
For liquid-cooled films, the competition between the bulk relaxation time 
$\tau_\alpha^b$
and the cooling rate $\gamma_c$ controls the distance to equilibrium.
This may be 
captured by the scaling relation 
\begin{equation}
U(\gamma_c,T) / U_e(T) = \mathcal{C}(x_c),
\label{eq:bulk}
\end{equation}
where $x_c = \gamma_c \tau_\alpha^b(T)/T$ represents the 
(adimensional) ratio between cooling and bulk relaxation timescales.
To establish that vapor-deposited films are controlled 
by the competition between the equilibrium 
relaxation time at the film surface $\tau_\alpha^s$
and the deposition rate $\gamma_d$ we seek a scaling relation of 
the same form, 
\begin{equation}
U(\gamma_d,T) / U_e(T) = \mathcal{D}(x_d),
\label{eq:surface}
\end{equation} 
where $x_d = \gamma_d \tau_\alpha^s(T)/\sigma_0$
is the (adimensional) ratio between deposition and surface relaxation timescales 
{and $\sigma_0$ is the average particle size}. {By construction,} the scaling
functions $\mathcal{C}(x)$ and $\mathcal{D}(x)$ 
should {both} converge to unity when $x \to 0$. 

\begin{figure}
\includegraphics[width=8.5cm]{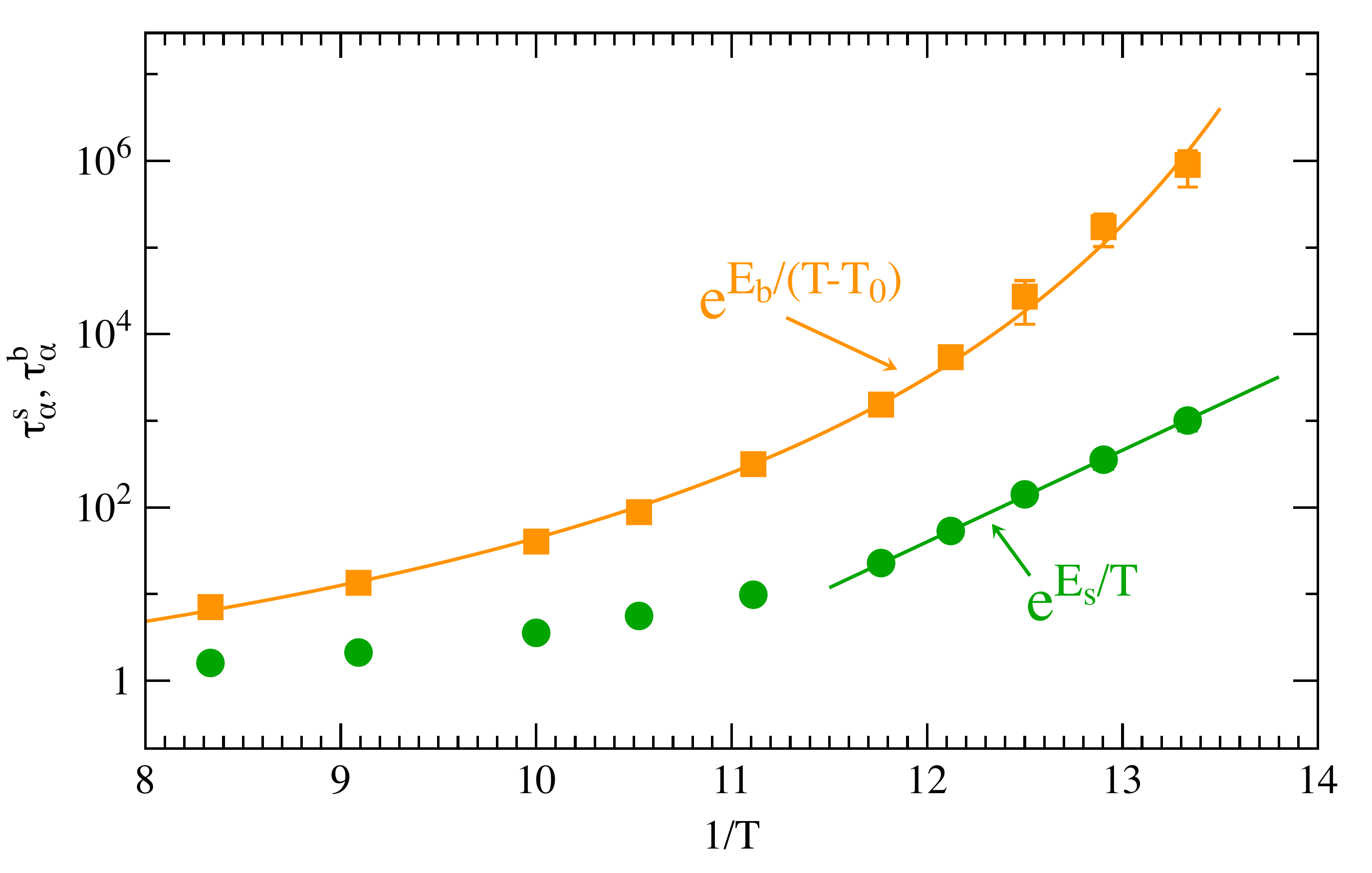}
\caption{\label{tau}
The relaxation times for the bulk region $\tau_\alpha^b$ (orange squares) and
the surface region $\tau_\alpha^s$ (green circles) both grow with decreasing temperature. The corresponding lines are fits to a Vogel-Fulcher form, 
$\tau_\alpha^b = \tau_0^b e^{E_b/(T-T_0)}$ with $T_0 = 0.0612 \pm 0.001$,
and an Arrhenius form, $\tau_\alpha^s = \tau_0^s e^{E_s/T}$.} 
\end{figure}
 
To test these scalings, we measure
the temperature dependence of the
{\it equilibrium} relaxation times at the film surface and in its core. 
To this end, we consider 
equilibrium films obtained from the swap Monte Carlo algorithm, 
that we then simulate using ordinary molecular dynamics.
The measurements described in the following
are thus generic for all films and are not specific to vapor-deposited ones.

Relaxation times are
extracted from the {decay of the} self-intermediate scattering function 
$F_s^l(q;t) = (1/N_l) \sum_{n=1}^{N_l} e^{i \mathbf{q} \cdot [\mathbf{r}_n(t) 
- \mathbf{r}_n(0)]}$,
{where the sum is restricted to the $N_l$ particles with positions 
$\mathbf{r}$ either in 
the bulk, $l=b$, or at the surface, $l=s$, of the film at time $t=0$.}
The chosen wavevector, $|\mathbf{q}| = 7.1$, coincides with the 
location of the first peak of the static structure factor, and is taken 
parallel to the surface. The bulk 
region is defined to be $5 \sigma_0$ thick at the
core of a film approximately $30 \sigma_0$ thick, while the two surface regions 
extend $1.5 \sigma_0$ from the film edge, as illustrated 
in the inset of Fig.~\ref{fs}. 

In Fig.~\ref{fs}(a), 
$F_s^b(\mathbf{q};t)$ appears typical of standard
glassy dynamics~\cite{Berthier2011}. 
A plateau emerges for $T \le 0.1$, and both its
length and height increase with decreasing temperature.
Particles are thus localized over an increasingly 
longer timescale upon increased cooling.   
The $F_s^s(\mathbf{q};t)$ in Fig.~\ref{fs}(b) are markedly different. 
First, there is no distinct plateau at any temperature. Second, a fast 
initial decay
down to $\sim0.5$ is followed by a slower relaxation. 
This suggests that surface particles perform
vibrational motion with a larger amplitude than {bulk ones}. 
More importantly, the long-time dynamics
at the surface is much faster than in the core. 

\begin{figure}
\includegraphics[width=8.5cm]{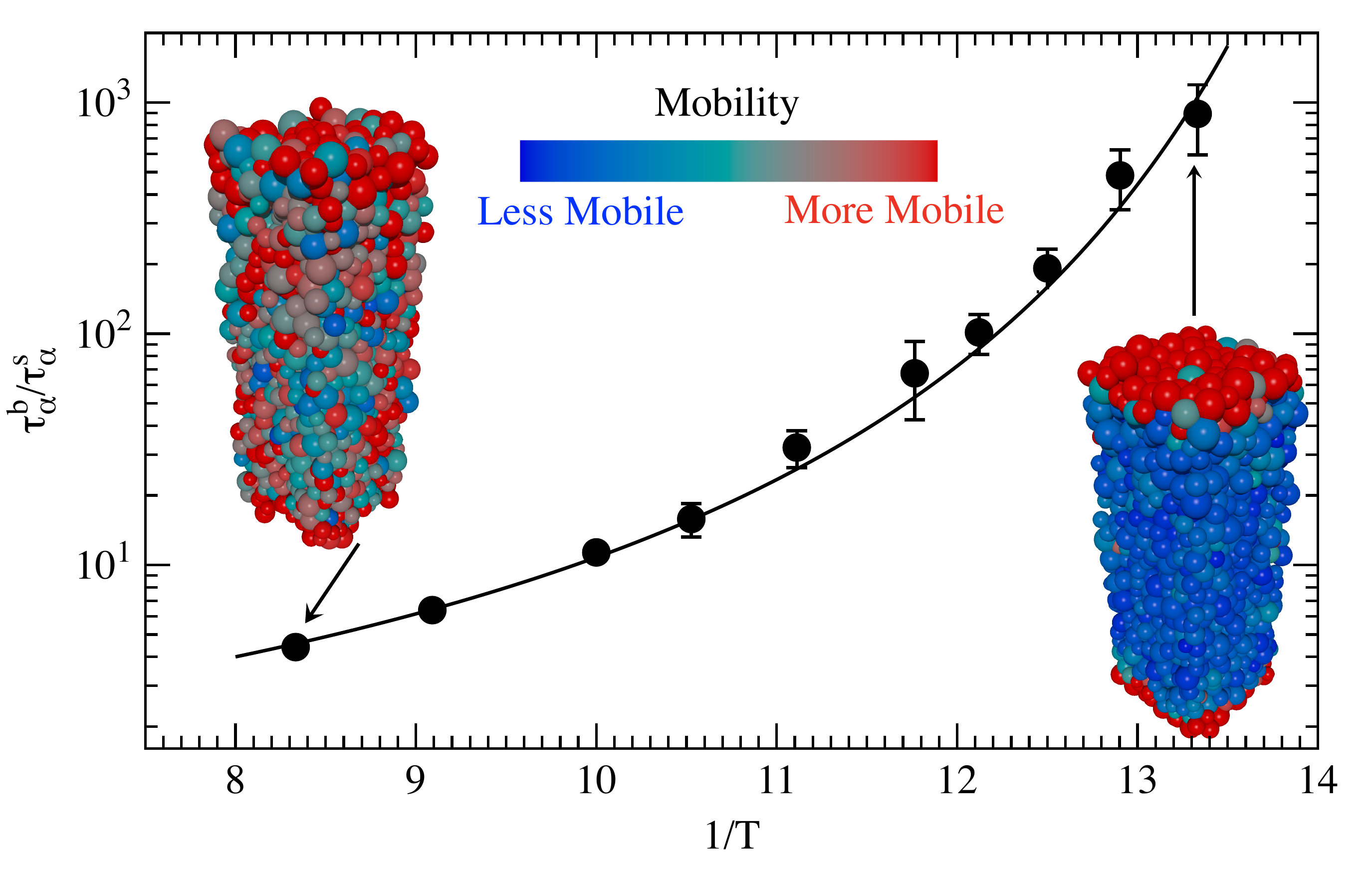}
\caption{\label{ratio}
The ratio $\tau_\alpha^b/\tau_\alpha^s$ grows rapidly with decreasing temperature. 
The line is an empirical fit to a Vogel-Fulcher form, 
$\tau_\alpha^b/\tau_\alpha^s \sim e^{E_r/(T-T_0)}$,
with the same $T_0$ as for $\tau_\alpha^b$. The 
insets show particles at time $t=0$ colored according to their 
displacement $20 \tau_\alpha^s$ later. Red particles have moved  
more than $\sigma_0$ and blue particles less than 
$0.5 \sigma_0$. 
The left inset shows that at $T=0.12$ particles within the core of 
the film move significantly over this timescale. The
right inset shows the emergence of a very thin mobile surface layer 
at $T=0.075$.}
\end{figure}

Defining the relaxation times
as $F_s^l(q;\tau_\alpha^l) = 0.2$
gives a measure of the time needed for a particle 
to move a distance comparable to its diameter. Empirically, 
$\tau_\alpha^b$ is well-described over the whole temperature range by 
a Vogel-Fulcher form $\tau_\alpha^b = \tau_0^b e^{E_b/(T - T_0)}$, 
with $T_0 = 0.0612 \pm 0.001$ (Fig.~\ref{tau}),
but other fitting forms {work equally well}. The surface relaxation time 
$\tau_\alpha^s$ also increases with decreasing temperature, 
although much less than $\tau_\alpha^b$,
and a Vogel-Fulcher form does not fit {it} 
well. Instead, $\tau_\alpha^s$ is better described by an 
Arrhenius {form}, 
$\tau_\alpha^s = \tau_0^s e^{E_s/T}$, at low 
temperatures, which is reminiscent of
the behavior of surface diffusion observed
experimentally~\cite{Yang2010,Zhu2011,Daley2012}. As a result, 
the ratio $\tau_\alpha^b/\tau_\alpha^s$ increases dramatically upon 
supercooling, growing from $\sim 4$ near the onset of localization 
to $\sim 900$ at the lowest temperature studied (Fig.~\ref{ratio}). 
Under the assumption (proved shortly below)
that the surface relaxation time controls the 
thermalization of the vapor-deposited glass, then the much faster 
surface relaxation is qualitatively consistent 
with the {enhanced thermalization} 
efficiency of vapor deposition observed 
in Fig.~\ref{energy}.

\begin{figure}
\includegraphics[width=3.2in]{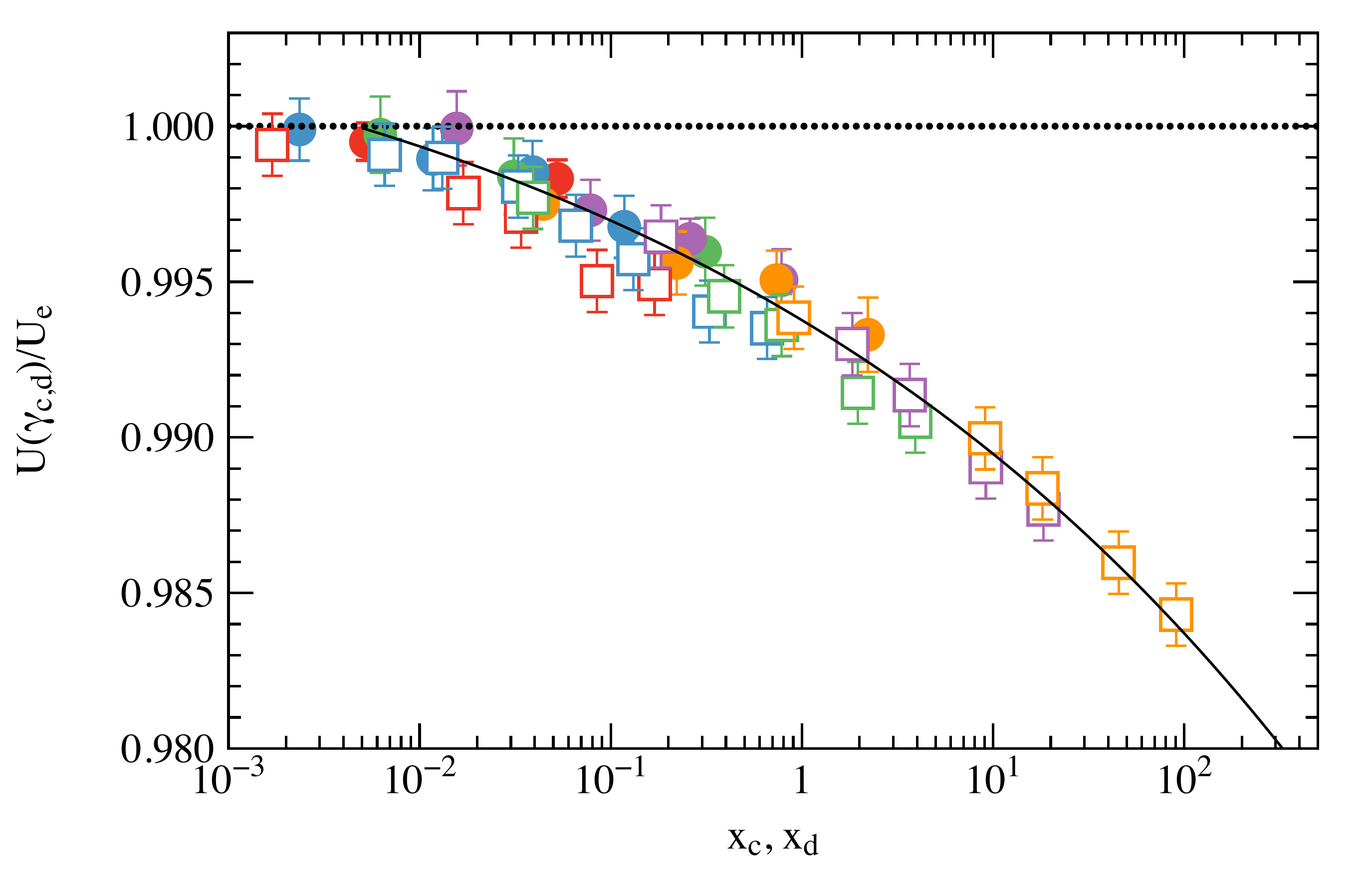}
\caption{\label{scaling}
The excellent collapse  observed for 
$U(\gamma_{c,d},T)/U_e$ as a function of $x_c = \gamma_c \tau_\alpha^b/T$ 
and $x_d = \gamma_d \tau_\alpha^s/\sigma_0$ for liquid-cooled films 
(squares), and vapor-deposited films (circles) 
indicates that surface relaxation plays the same role in the formation of 
vapor-deposited films as bulk relaxation in 
ordinary liquid-cooled film formation. Results are reported for
temperatures $T=0.085$ (red), $T=0.0825$ (blue), $T=0.08$ (green), 
$T=0.0775$ (purple), and $T=0.075$ (orange). The solid line is 
an empirical fit.}
\end{figure}

To visualize the emergence of a mobile surface layer, 
we color particles according to their 
displacement $| \mathbf{r}_n(t) - \mathbf{r}_n(0) |$
after $t = 20 \tau_\alpha^s$ for 
the highest and lowest temperature examined as insets to Fig.~\ref{ratio}. 
At all temperatures the surface is more mobile than the core, {but qualitative differences can be observed. At the highest temperature mobile particles 
are found throughout the film, while at the lowest temperature 
only a small layer of mobile particles is observed. This layer is barely thicker 
than $\sigma_0$ in the inset of Fig.~\ref{ratio}.
(Fitting an exponential decay to the surface relaxation time 
gradient, see Ref.~\cite{starr2017}, gives a thickness of order 2-3 $\sigma_0$ with a weak temperature dependence.)
Such a decoupling between bulk and surface
dynamics has been experimentally documented~\cite{zara2017}, 
but was not directly connected to the thermalization of ultrastable glasses before.

The mobile surface layer is exploited by the vapor-deposition 
process to speedup
the thermalization of glassy films, as we can directly demonstrate.
{The combination in Fig.~\ref{scaling} of all our energy measurements 
in liquid-cooled and vapor-deposited films 
shows that Eqs.~(\ref{eq:bulk}, \ref{eq:surface}) collapse
the numerical results very well.}
Moreover, the simulations indicate that 
the scaling functions $\mathcal{C}(x)$ and $\mathcal{D}(x)$
are nearly identical. Note that
no adjustable parameter is used for these scalings, which combine 
independent numerical measurements. 

The excellent data collapse in Fig.~\ref{scaling} indicates that the 
surface relaxation time and deposition rate 
determine the distance to equilibrium for vapor-deposited films
in the exact same way that 
bulk relaxation time and cooling rate control the distance 
from equilibrium for liquid-cooled films. 
This result suggests that one can convert the deposition rate of a film into an effective cooling rate as 
$\gamma_c^{\rm eff} = \gamma_d (\tau_\alpha^s/\tau_\alpha^b) (T/\sigma_0)$. 
Quantitatively, we can fit 
the scaled data to an empirical power law, 
$\mathcal{P}(x) = a x^{\nu} + b$ (with $\nu=0.12 \pm 0.03$, solid line
in Fig.~\ref{scaling}) \cite{Reid2016}. 
Solving for $\mathcal{P}(x) = 1$ gives the maximal rate, for both preparation processes, at which 
equilibrium films can be prepared. 
Our simulations thus provide a simple quantitative criterion,
$x_d = \gamma_d \tau_\alpha^s/\sigma_0 \leq 0.005$, 
to create equilibrium vapor-deposited 
films. 

Consistency with experiments is illustrated by  
considering the case of indomethacin~\cite{Swallen2007,Zhu2011}, for which
we use $\sigma_0=1$~nm~\cite{Zhu2011} as the length unit and $10^{-12}$~s as the time 
unit~\cite{Xiang2013}. 
We further approximate the structural relaxation times using 
$\tau_\alpha^l = (q^2 D^{l})^{-1}$ at $q = 2 \pi/\sigma_0$ 
with the diffusion coefficients $D$ obtained in Ref.~\cite{Zhu2011}. 
For a cooling rate of
$40$~K/min, Swallen \textit{et al.}\ determined that $T_g = 315$~K \cite{Swallen2007}.
We thus estimate that $\gamma_c \tau_\alpha^{b}(T_g)/T_g \approx 0.0054$, which is in close agreement with the above criterion. Our proposed scaling form, thus captures well the cooling rate 
dependence of the glass transition. More interestingly, we can compute the lowest temperature at which one can obtain an 
equilibrium film by vapor deposition for a given deposition rate. 
For $\gamma_d = 0.2$~nm/s \cite{Dalal2013}, 
we estimate that the smallest surface diffusion coefficient at which an equilibrium film
can be obtained should be $D^s \approx 1.0 \times 10^{-18}$ m$^2$/s.
By extrapolating the surface diffusion coefficients in Ref.~\cite[Fig.~3]{Dalal2013},
we find that an equilibrium vapor-deposited film 
should be accessible 
down to $264~{\rm K} \approx  0.84 T_g$, which is close
to the experimental estimate \cite{Dalal2013}. 
Note that for simulations 
our results imply that an efficiency gain of at most 2-3 orders of magnitude 
can be expected from vapor deposition over standard bulk annealing, which appears consistent with a recent independent estimate~\cite{starr2017}. 
In the regime accessible to experiments, by contrast, that gain can reach eight orders of magnitude.   

Our work directly and quantitatively 
demonstrates how enhanced surface diffusion, 
as quantified by $\tau_\alpha^s$, is responsible for the ultrastability
of vapor-deposited films. Additional considerations 
may be needed to account {for the full scope of experimental observations}. 
First, additional studies are needed to connect the surface relaxation time 
measured in this study with the surface mobility inferred from experiments 
\cite{Rodriguez2016,Swallen2007,Kearns2010,Zhu2011,Daley2012,zara2017,Malshe2011}.
Second, the shape and chemical nature of 
vapor-deposited molecules can result in preferential 
orientation within the vapor-deposited 
film \cite{Lyubimov2015,Dalal2012a,Yokoyama2009,Yokoyama2010,Dalal2015,Jiang2016}, 
while such alignment bias is not expected for ordinary 
liquid-cooled films. Although
molecular alignment can be used to tailor glassy properties \cite{Dalal2015,Jiang2016}, it 
also inherently leads to vapor-deposited films that differ {in structure} from their liquid-cooled counterpart. 
For the simple, spherical particles studied in this work, however, we find no evidence in the pair-correlation function (not shown) or the density profile that vapor deposition produces different structures or particle segregation profiles
than what is seen in liquid cooling. Our results are thus consistent with those of Reid \textit{et al.}\ \cite{Reid2016}
for a binary Lennard-Jones system. 
Overall, this work is a first step to obtain a more quantitative understanding of the 
creation of vapor deposited supercooled liquids and glasses, and additional work is needed to understand the
role of the substrate, molecular shape, and other factors in this process. 

\begin{acknowledgments}
E.F. acknowledges funding from NSF DMR-1608086. 
This work was supported by a grant from the Simons 
Foundation (\#454933 Ludovic Berthier,
\#454937 Patrick Charbonneau, 
\#454955 Francesco Zamponi).  
We thank Mark Ediger, Beatriz Seoane, and Grzegorz Szamel 
for many useful discussions. 
Data relevant to this work have been archived and can be accessed at \url{https://doi.org/10.7924/G8P26W5G}
\end{acknowledgments}

\end{document}